# Field-driven skyrmion motion through velocity equipartition between skyrmions and a domain wall

Xiangjun Xing[1,2,*] and Yan Zhou[3,*]

[1]School of Physics & Optoelectronic Engineering, Guangdong University of Technology, Guangzhou 510006, China

[2]Guangdong Provincial Key Laboratory of Information Photonics Technology, Guangdong University of Technology, Guangzhou 510006, China

[3]School of Science & Engineering, The Chinese University of Hong Kong, Shenzhen, Guangdong 518172, China

**ABSTRACT**

Magnetic skyrmions, as a whirling spin texture with axisymmetry, cannot be propelled by a uniform magnetic field. Therefore, reported skyrmion motions have been induced using other sorts of stimuli; typically, electric currents in magnetic metals as well as spin waves, thermal gradient, and field gradient have manifested their ability to drive skyrmion motion. Here, we demonstrate, through micromagnetic simulations and analytically, that magnetic skyrmions can be displaced by a uniform perpendicular magnetic field via a domain-wall (DW) mediator. For a fixed field strength, the velocity of a skyrmion train evolves in terms of $1 / (1 + N_s)$ with $N_s$ denoting the number of skyrmions. Based on Thiele's model, we reproduce the velocity–$N_s$ relation first identified from numerical results and reveal that the skyrmion–DW and inter-skyrmion repulsions offer the direct driving force for skyrmion motion. This study underlines the role of spin textures' interaction in skyrmion dynamics, and opens an alternative route for skyrmion manipulation especially relevant to insulating magnets. Given the correspondence between magnetism and electricity, we anticipate that the scheme should also work for polar skyrmions in ferroelectrics.

[*]Email: xjxing@gdut.edu.cn; zhouyan@cuhk.edu.cn

**INTRODUCTION**

Since their experimental discovery, magnetic skyrmions have drawn unprecedented attention of global researchers for their importance from both fundamental and applied perspectives [1–3]. Now, it is confirmed that magnetic skyrmions can be stabilized by several mechanisms [1]. Because of the integer topological charge (also called skyrmion number), magnetic skyrmions exhibit emergent electrodynamics [4]: electrons flowing across magnetic skyrmions gain a Berry phase and thus undergo an excess transverse deflection, which is measurable as topological Hall resistance; conversely, magnetic skyrmions driven by the electrons' spin transfer travel at an angle with respect to the current flow, experiencing skyrmion Hall effect. The two conjugate effects can be used to probe and harness magnetic skyrmions coded as data bits in e.g., racetrack memory [5]. Actually, magnons impinging on magnetic skyrmions can induce similar outcome [6,7]. Furthermore, magnetic skyrmions can be driven by thermal or field gradient to rotate like a ratchet [8–11]. Despite a decade of intense research, there remain a gap and controversies between experiments and theory regarding magnetic skyrmion motion, such as the skyrmion-size dependency of skyrmion velocity. In real samples, interplay of skyrmions and pinning sites/neighboring magnetic textures may play a dominant role in skyrmion motion, and lead to deviation of skyrmion behaviors from theoretical anticipation [12–13].

For the common magnetic skyrmions, i.e., axisymmetric, on ferromagnetic background, and originating from Dzyaloshinskii-Moriya interaction (DMI) [14–15], the inter-skyrmion, skyrmion–boundary, and skyrmion–DW interactions are repulsive [16–18]. In reality, the rectilinear motion of magnetic skyrmions along a nanotrack is guaranteed by the skyrmion–boundary repulsion that cancels out the Magnus force [18,19]. Namely, these mutual interactions can influence skyrmion motion in a constructive or destructive manner. Mastering how to steer them will help avoid detrimental effect and may open new avenues for skyrmion manipulation.

Magnetic DWs ubiquitously exist in magnets required by tradeoff between short-range exchange and long-range dipolar interactions. In chiral magnets, innate or induced, magnetic skyrmions can subsist, due to DMI, as a metastable or even stable spin configuration [3]. Over a broad parameter space, magnetic skyrmions and DWs can coexist in the same magnet, especially for constricted geometry [12,20]. When harboring a DW, a given ferromagnet no longer exhibits mirror symmetry, and under a magnetic field, the DW will migrate to expand the domain parallel to the field [21]. Different from DWs, skyrmions, as localized topological solitons, embody revolution symmetry, and thus, in a perpendicular magnetic field, they will expand or contract without substantial displacement of their centers [3,20,22]. That is why no field-induced magnetic skyrmion motion is reported.

In a perpendicular field, a DW will inevitably approach the skyrmions. As a result, the DW and neighboring skyrmion must repel each other. Then, the skyrmion is supposed to move under action of the repulsive force. In this paper, using micromagnetic simulations and analytically, we corroborate the above conjecture based on a magnetic nanotrack, which allows for simultaneous realization of skyrmion–DW, inter-skyrmion, and skyrmion–boundary interactions. Our simulations demonstrate that, irrespective of the helicity, i.e., Néel or Bloch type, a train of skyrmions can be set into coherent motion alongside a DW by a uniform magnetic field, and noteworthily, their velocity drops down to 1 / (1 + $N_s$) with $N_s$ the number of included skyrmions. The analytical result derived from the Thiele equation agrees well with and explains the numerical finding. Skyrmion annihilation at the nanotrack edge is also seen at high magnetic fields, when the skyrmion–boundary repulsion cannot balance the Magnus force [18,19]. These results may find applications in skyrmion-based sensing devices. Apart from magnetic skyrmions, the proposed field-drive of skyrmion motion appears valid for polar skyrmions in ferroelectric materials [23,24], where an electric version of DMI has been identified theoretically through symmetry analysis [25]. This is a point worthy of experimental exploration as polar topological structures may indicate unique dynamics unattainable in their magnetic counterparts [26].

**MICROMAGNETIC MODEL AND SIMULATIONS**

This study concentrates on common skyrmions. To this end, DMI is incorporated in the free energy of the target system — a magnetic nanotrack, 60 nm wide and 1 nm thick. Perpendicular magnetic anisotropy (PMA) is also included in the model, allowing for stabilization of skyrmions without applying a field. An out-of-plane magnetic field serves as the driving force for skyrmion dynamics. Thus, the total free energy, $E$, comprises the contributions of exchange, dipolar, PMA, DMI and Zeeman interactions.

An initial, equilibrium magnetization configuration in the nanotrack is obtained by relaxing the entire system with a properly set near-equilibrium configuration until $E$ reaches a minimum. Subsequently, to trace the dynamics of the resulting equilibrium magnetization configuration, we numerically solve the Landau-Lifshitz-Gilbert equation [27,28],

$$\frac{\partial \mathbf{m}}{\partial t} = -\gamma \mathbf{m} \times \mathbf{H}_{\text{eff}} + \alpha \mathbf{m} \times \frac{\partial \mathbf{m}}{\partial t}, \qquad (1)$$

where $\mathbf{m} = \mathbf{M}(\mathbf{r}, t) / |\mathbf{M}|$ is continuous function of the coordinate $\mathbf{r}$ and time $t$, and $\mathbf{H}_{\text{eff}} = -\frac{1}{\mu_0 |\mathbf{M}|} \frac{\delta E}{\delta \mathbf{m}}$ represents the effective field with $\mu_0$ denoting the vacuum magnetic permeability; $\gamma$ and $\alpha$ stand for the gyromagnetic ratio and Gilbert damping constant, respectively.

Numerical simulations are implemented using the finite-difference code, MuMax3 [29]. In all simulations, the nanotrack is divided into a grid of $1 \times 1 \times 1$ nm$^3$ cubes, whose side length is small compared to the relevant feature sizes [30–32]. Exchange stiffness $A = 15$ pJm$^{-1}$, saturation magnetization $M_s$ = 580 kAm$^{-1}$, and Gilbert damping constant $\alpha$ = 0.3 are fixed, whereas various values of PMA constant $K_u$ and DMI strength $D$ are addressed individually. Unless otherwise specified, the results presented are based on $K_u$ = 0.8 MJm$^{-3}$ and $D$ = 3.5 mJm$^{-2}$. We carry out two sets of simulations, one with interfacial DMI and the other with bulk DMI, accommodating Néel and Bloch skyrmions, respectively. In principle, magnetic materials with $C_{nv}$ ($T$) symmetry permit occurrence of the interfacial (bulk) DMI [33]. Open boundary conditions are assumed.

**RESULTS**

**An isolated skyrmion in magnetic field**

In B20 magnets, the A phase, i.e., the skyrmion lattice state, emerges at a finite magnetic field [1,33], which is equivalent to a PMA. The equilibrium size of skyrmions in this phase is a function of applied magnetic field [3]. Later, it is seen that the size of isolated skyrmions in ultrathin multilayer films with an actual PMA also sensitively depends on the applied field [20,22]. Here, we numerically explore how the magnetic field affects an isolated skyrmion. The results are presented in Fig. 1. Note that, in this study, we focus on the steady-state behaviors of magnetic textures and ignore the initiation process approaching equilibrium. Hence, all results presented correspond to certain equilibrium states, either static or dynamic.

Figure 1(a) shows the skyrmion evolution with applied fields in the narrow nanotrack (60 nm wide). Very clearly, the skyrmion adapts itself to the magnetic field. For a positive field, the skyrmion expands with the field strength, and for a negative field, it shrinks with the field strength. This skyrmion size variation stems from the energy redistribution in the system. At the highest field, the circular skyrmion is elongated in the longitudinal direction, because of the transverse confinement imposed by the edges [34].

For comparison, we also examine the skyrmion evolution in an extended area; as shown in Fig. 1(b), where a 300 nm wide square element is used. At first glance, there is no remarkable difference from the above situation. The field dependency of the skyrmion diameter follows the same tendency. However, the geometric confinement has a significant impact on field susceptibilities of the skyrmion size [Fig. 1(c)] and shape [Fig. 1(a), bottom panel & Fig. 1(b), rightmost panel]. Effectively, the role of confined geometry resembles a magnetic field opposite to the skyrmion core polarity. In the square element, the skyrmion grows rapidly and turns to a big bubble suddenly at a relatively small field; whereas in the nanotrack, the skyrmion grows slowly and saturates till very large fields. Overall, these findings are in accordance with reported results, and no translational motion of skyrmion is observed.

**A domain wall in magnetic field**

After decades of research, the DW dynamics has been well comprehended. It is generally accepted that, at small fields, the DW travels at a speed directly proportional to the field strength, and above a threshold field, the DW velocity drops abruptly due to Walker breakdown [31,35,36]. Recently, it is further recognized that the chiral DW induced by DMI tilts during motion and the tilting modifies DW dynamics [37].

We simulate DW motion in a magnetic field and derive its velocity along the nanotrack. The DW's snapshots, at several particular times after applying the field, are shown in Fig. 2(a). As expected, DW tilting occurs. After entering steady state, the DW proceeds at a uniform speed (203 nm per 1.7 ns in this case) with a fixed tilt angle.

The DW velocity versus field curves, for a series of ($K_u$, $D$) combinations, are plotted in Fig. 2(c) [38], from which several signatures can be identified. Prominently, all curves rise monotonically with the field. After scrutiny, one can find slight nonlinearity on the curves. This nonlinear deviation stems from the DW tilting, as elucidated in Ref. [37]. For different ($K_u$, $D$), the curves distinguish from each other by the slope. These results agree with the literature [37].

Above a certain threshold field (see supplemental Table S1 [39]), the Walker breakdown in DW motion occurs periodically, which is clearly reflected in supplemental Fig. S1(a), where the total topological charge of the system oscillates with time around a value of 0.5 characteristic of an antivortex [40]. In this process, an antivortex-like spin texture is firstly nucleated [31,41] at and then divided from an end of the DW and ultimately annihilated around the sample edge, as shown in Fig. S1(b).

**A domain wall–skyrmion ensemble in magnetic field**

To step further, we insert a skyrmion behind the DW in the nanotrack, which is then immersed in a perpendicular magnetic field. The field-induced dynamics of the DW–skyrmion pair is depicted in Fig. 2(b). It is seen that the skyrmion moves together with the DW. Traversing a 209 nm interval, the pair spends 3.4 ns, resulting in a velocity 61.5 m/s at 400 Oe. Therefore, in a uniform magnetic field, the DW and accompanying skyrmion can act cooperatively and move as a whole. It is easy to find that such synergetic motion does not hold for a DW–DW pair and a skyrmion–skyrmion pair, which will move oppositely and remain stationary, respectively, in response to the field.

By contrast, without the skyrmion, the single DW takes only 1.7 ns through 203 nm [see Fig. 2(a)], giving a velocity 119.4 m/s. That is, inclusion of the skyrmion slows down the DW by a factor of ~ 1/2. For a systematic comparison, we simulate the DW–skyrmion pair's motion at various ($K_u$, $D$) against the field strength. The velocity–field curves are plotted in Fig. 2(c). It is apparent that, the pair is always slower than the single DW. After a careful inspection, one finds that the reduction in the DW velocity is ~ 1/2.

When the field is large enough, the skyrmion will touch the sample edge and be annihilated, as illustrated in Fig. S2(b). At the point of the skyrmion annihilation, the total topological charge of the system jumps from 1 to 0, as shown in Fig. S2(a). The reason for the skyrmion annihilation is that, at the suprathreshold fields (see Table S1), the edge repulsive force on the skyrmion can no longer offset the Magnus force that scales linearly with the skyrmion velocity [18,19]. Indeed, as long as the skyrmion is sufficiently fast, the Magnus force will necessarily surpass the edge confining force resulting in the skyrmion annihilation, regardless of the driving scheme.

In the following, we place more skyrmions behind the DW and see how this ensemble behaves in the perpendicular field. The velocity–field curves, for the ensemble hosting 1–5 skyrmions, are shown in Fig. 3(b), and exemplarily, the images of the ensemble with 5 skyrmions, at particular times after application of the field, are shown in Fig. 3(a).

It is observed that the ensemble can still be moved forward by the external field when more than 1 skyrmions participate. For instance, in the 5-skyrmion case, the ensemble travels 174 nm in every 8 ns at 400 Oe [Fig. 3(a)]. The velocity of the ensemble linearly increases with the applied field but falls further as the number of the skyrmions increases [Fig. 3(b)]. Another striking feature is that, during motion, individual skyrmions in the ensemble do not have identical sizes, and the skyrmions farther from the DW turn out to be larger, as shown in Fig. 3(a). To unveil more details, we derive the DW–skyrmion and inter-skyrmion separations as well as the diameter of each skyrmion; the results for the 5-skyrmion case are shown in Fig. 3(c). It indicates that an elevation in the field strength leads to a bigger spreading of the skyrmion sizes and a smaller separation between neighboring solitons. This trend holds true universally for all cases with multiple skyrmions.

To be more informative, the velocity–field curves in Fig. 3(b) are replotted as the velocity–$N_s$

curves, as shown in Fig. 4(a). Apparently, for all field values, the ensemble velocities decrease with the number of skyrmions in seemingly the same manner. Furthermore, we define the mobility of an ensemble $u$ as $u = \frac{dv}{dH}$ [36], where $v$ and $H$ are the velocity of the ensemble and external field, respectively. In this context, each curve in Fig. 3(b) gives rise to a mobility value, which is in turn plotted in Fig. 4(b) against $N_s$, i.e., the number of skyrmions included in the ensemble. In this figure, the mobility values, $\tilde{u}$, of the ensemble are well fitted by the formula,

$$\tilde{u} = \tilde{u}_{dw} / (N_s + 1), \qquad (2)$$

where $\tilde{u}_{dw}$ is the mobility value of a single DW carrying no skyrmions. That is to say, at a given field, the ensemble velocity is inversely proportional to the total number of the solitons (i.e., DW and skyrmions) in the ensemble. The more the skyrmions are included, the slower the ensemble moves. When skyrmions are transported, the original DW velocity is equally distributed amongst all solitons in the ensemble.

**Thiele model of ensemble motion**

Projecting Eq. (1) onto the ensemble translation mode under the assumption that each spin texture in the ensemble owns a rigid structure, i.e., $\mathbf{m}(\boldsymbol{r}, t) = \mathbf{m}[\boldsymbol{r} - \boldsymbol{\mathcal{R}}(t)]$ and moves steadily, i.e., $\boldsymbol{\mathcal{R}}(t) = \mathbf{v}t$ (where $\boldsymbol{\mathcal{R}}(t)$ and $\mathbf{v}$ are the center of mass and velocity of a spin texture, respectively), one obtains the Thiele equation [31,42],

$$\mathbf{F}^G - \mathbf{F}^\mathcal{D} + \mathbf{F} = 0, \qquad (3)$$

where $\mathbf{F}^G$ and $\mathbf{F}^\mathcal{D}$ are the Magnus force and dissipative force, respectively, and $\mathbf{F}$ represents a force due to the external field, inter-soliton repulsion, edge confinement, or impurities. Eq. (3) dictates that all the forces on a steadily moving spin texture are in balance.

In this study, we concentrate on the motion of the ensemble along a nanotrack, where the soliton transverse motion is suppressed by the edge confinement, i.e., $\mathbf{v} = (v, 0)$. Hence, $\mathbf{F}^{\mathcal{D}} = \alpha \overleftrightarrow{\mathcal{D}} \cdot \mathbf{v} = \alpha \begin{pmatrix} \mathcal{D}_{xx}, \mathcal{D}_{xy} \\ \mathcal{D}_{yx}, \mathcal{D}_{yy} \end{pmatrix} \begin{pmatrix} v \\ 0 \end{pmatrix} = \alpha \begin{pmatrix} \mathcal{D}_{xx} \\ \mathcal{D}_{yx} \end{pmatrix} v$, where $\overleftrightarrow{\mathcal{D}} = \begin{pmatrix} \mathcal{D}_{xx}, \mathcal{D}_{xy} \\ \mathcal{D}_{yx}, \mathcal{D}_{yy} \end{pmatrix}$ is the dissipation dyadic of a spin texture [31,42]. Based on this fact, the transverse forces are disregarded in the following analysis.

To elucidate the numerical findings, Eq. (3) is applied to every soliton in the ensemble (as shown in Fig. 5), and considering solely the longitudinal forces, one has,

$$\begin{cases} F_{dw}^{\mathcal{D}} + F_{10} = F^H, & \text{DW}, \\ F_{sk1}^{\mathcal{D}} + F_{21} = F_{01}, & \text{SK1}, \\ F_{sk2}^{\mathcal{D}} + F_{32} = F_{12}, & \text{SK2}, \\ F_{sk3}^{\mathcal{D}} = F_{23}, & \text{SK3}, \end{cases} \quad (4)$$

where $F_{dw}^{\mathcal{D}}$ and $F_{ski}^{\mathcal{D}}$ are the dissipative forces on the DW and $i$th skyrmion, respectively, $F_{i+1,i}$ and $F_{i,i+1}$ are the mutual repulsive forces between two neighboring solitons (note that, for the DW $i = 0$; for the skyrmions $i = 1, 2, 3, \ldots N_s$), and $F^H$ stands for the driving force on the DW exerted by the external field.

Adding together all the equations in Eq. (4), one gets,

$$F_{dw}^{\mathcal{D}} + \sum_1^3 F_{ski}^{\mathcal{D}} + \sum_0^2 F_{i+1,i} = F^H + \sum_0^2 F_{i,i+1}. \quad (5)$$

Provided that there are totally $N_s$ skyrmions, it is easy to find,

$$F_{dw}^{\mathcal{D}} + \sum_1^{N_s} F_{ski}^{\mathcal{D}} + \sum_0^{N_s-1} F_{i+1,i} = F^H + \sum_0^{N_s-1} F_{i,i+1}. \quad (6)$$

According to the Newton's third law, $F_{i+1,i} = F_{i,i+1}$, since they are a pair of action and reaction forces. Then, Eq. (6) becomes,

$$F_{dw}^{\mathcal{D}} + \sum_1^{N_s} F_{ski}^{\mathcal{D}} = F^H. \quad (7)$$

For an out-of-plane field, the Zeeman energy of the system is $E_z = \mu_0 M_s H d w (l - 2q) - N_s \mu_0 M_s H d \mathcal{S}$, where $d$, $w$, and $l$ denote the thickness, width, and length of the nanotrack, respectively, $q$ is the position of the DW center, and $\mathcal{S}$ is the area of a skyrmion. Correspondingly, by definition, $F^H = -\frac{\partial E_z}{\partial x} = 2\mu_0 M_s H d w$ [31].

In the case of steady ensemble motion, $v_{\mathrm{dw}} = v_{\mathrm{ski}} = v$, and specifically, $F_{\mathrm{dw}}^{\mathcal{D}} = \alpha \mathcal{D}_{xx}^{\mathrm{dw}} v$ and $F_{\mathrm{ski}}^{\mathcal{D}} = \alpha \mathcal{D}_{xx}^{\mathrm{ski}} v$. To express $\mathcal{D}_{xx}^{\mathrm{dw}}$ and $\mathcal{D}_{xx}^{\mathrm{ski}}$ explicitly, we choose the following ansatzes for the DW [37] and skyrmion [22], respectively,

$$\theta(x,y) = 2\arctan\left(e^{\frac{(x-q)\cos\chi - y\sin\chi}{\Delta}}\right), \qquad (8)$$

$$\text{and } \theta(r) = \arcsin\left[\tanh\left(\frac{r+R}{\Delta/2}\right)\right] + \arcsin\left[\tanh\left(\frac{r-R}{\Delta/2}\right)\right], \qquad (9)$$

where $\theta$ and $\phi$ are the polar and azimuthal angles of magnetization with $\mathbf{m} = (\sin\theta\cos\phi, \sin\theta\sin\phi, \cos\theta)$, $x$ and $y$ are the Cartesian coordinates along the length and width of the nanotrack, and $\chi$ and $\Delta$ are the tilt angle and thickness of the DW, respectively; $r$ is the radial distance relative to the skyrmion center at $(x,y)$ and $R$ is the skyrmion radius.

For the DW, $\mathcal{D}_{xx} = \frac{\mu_0 M_s d}{\gamma}\iint\left(\frac{\partial\theta}{\partial x}\right)^2 dxdy$, substituting Eq. (8) into which, one obtains $\mathcal{D}_{xx}^{\mathrm{dw}} = 2\frac{\mu_0 M_s d}{\gamma}\frac{w}{\Delta}$. For a skyrmion, $\mathcal{D}_{xx} = \frac{\pi\mu_0 M_s d}{\gamma}\int\left[\left(\frac{d\theta}{dr}\right)^2 + \left(\frac{\sin\theta}{r}\right)^2\right]rdr$, substituting Eq. (9) into which, one arrives at $\mathcal{D}_{xx}^{\mathrm{ski}} \approx 2\pi\frac{\mu_0 M_s d}{\gamma}\frac{R}{\Delta}$ [43].

As already noted, during the steady motion, different skyrmions in the ensemble do not have identical sizes. The ones nearer to the DW are smaller because of the increasingly larger repulsive forces. For the $i$th skyrmion away from the DW, the radius $R_i \approx R_1 + (i-1)\delta$ with $\delta$ denoting the difference in radii of two neighboring skyrmions. Then, one finds $\sum_1^{N_s}\mathcal{D}_{xx}^{\mathrm{ski}} = 2\pi\frac{\mu_0 M_s d}{\gamma}\left[N_s\frac{R_1}{\Delta} + \frac{N_s(N_s-1)}{2}\frac{\delta}{\Delta}\right]$ and thus the ensemble velocity reads,

$$v = \frac{\frac{\gamma}{\alpha}H\Delta}{\eta N_s + 1} \approx \frac{v_{\mathrm{dw}}}{N_s+1}, \qquad (10)$$

where $v_{\mathrm{dw}} = \frac{\gamma}{\alpha}H\Delta$ is the velocity of a single DW in a field $H$ [31] and according to the simulation results $\eta = \pi\frac{R_1 + \frac{N_s-1}{2}\delta}{w} \approx 1$.

From Eq. (10), it is obvious that the ensemble mobility $u \approx \frac{u_{\mathrm{dw}}}{N_s+1}$ (wherein $u_{\mathrm{dw}} = \frac{\gamma}{\alpha}\Delta$ represents the mobility of a single DW), which perfectly replicates the empirical formula [Eq. (2)] derived from numerical results.

**DISCUSSION**

The Thiele model captures the essence of field-driven dynamics of the DW–skyrmion ensemble, in which the repulsive forces between neighboring solitons efficiently drives the motion of the skyrmions but as internal forces do not enter the final equation, i.e., Eq. (7), enabling the derivation of the ensemble velocity without knowing specific forms of the repulsive forces [16,18]. The equipartition of the original DW velocity among the DW and skyrmions has its roots in the distribution of the external force experienced by the DW over all the solitons in the ensemble via the DW–skyrmion and inter-skyrmion interplays, as schematically illustrated in Fig. 5. Moreover, Eq. (4) reveals that the repulsive force acting on the $j$th skyrmion by the left-hand soliton is $F_j = \sum_j^{N_s} F_{ski}^{\mathcal{D}}$, namely, the nearer the skyrmion to the DW, the larger the stress experienced by the skyrmion (see Fig. 5 for a 3-skyrmion instance), which underlies why the skyrmion closer to the DW has a smaller size, as shown in Figs. 3(a), S3(a), and S4(a). Actually, this scenario resembles skyrmion compression in a nanotrack with enhancing geometric confinement [2,19].

As seen from Figs. 2(c) and 3(b), attaching skyrmions to the DW can eliminate the nonlinearity on the velocity–field curves for the single DW, which is ascribable to the greatly reduced tilt angle of the DW in the presence of skyrmions [37]. During motion, the DW tends to tilt with respective to the transverse direction and the skyrmion deflects to an edge of the nanotrack. The directions of the DW tilting and skyrmion deflection match each other in such a way that the DW–skyrmion repulsion suppresses the DW tilting.

The demonstrated scheme of field-induced skyrmion motion is validated in a broad region of parameter space, as already shown in Fig. 2(c). The data for ensemble motion under different sets of ($K_u$, $D$) as presented in Fig. S3 solidifies the notion. Furthermore, for the Bloch-type spin textures, the drive scheme also works well, as shown in Fig. S4. In fact, this point is verified by analytic results based on the Thiele model, since in both the DW and skyrmion ansatzes the azimuthal angles $\phi$ describing the DW and skyrmion helicities are not explicitly specified [see Eqs. (8) and (9)]. These results suggest that the proposed field-based drive of skyrmion motion is universally applicable to common magnetic skyrmions stabilized by DMI.

In a configuration composed of a DW and two sets of skyrmions with oppositely oriented cores, one can selectively transfer either set of the skyrmions along a certain direction by toggling the field polarity, while leaving the other set undisturbed, as shown in Fig. 6. Such operation is impractical in the current-driven case where all skyrmions, whatever the core polarities, tend to shift at the same pace. This behavior may enable the realization of field-controlled unidirectional devices.

Finally, we would like to discuss the relevance of these results to polar skyrmions [23,24] in ferroelectric materials. Up to now, electric/spin currents, spin waves, and temperature and field gradients have proved capable of displacing magnetic skyrmions [4,6–13]. Amongst them, the most fascinating is the current-induced skyrmion motion, which holds promise in building compact, energy efficient skyrmionic devices [2]. The field-drive of skyrmion motion conceived in this study represents an alternative route awaiting experimental exploration and might find use in magnetic insulators.

As the counterpart of magnetic skyrmions, polar skyrmions have been experimentally observed in both multilayered ($PbTiO_3$/$SrTiO_3$) [23] and bulk ($Bi_{0.5}Na_{0.5}TiO_3$) [24] ferroelectrics. Theoretical strategies to robustly imprint skyrmions in simple ferroelectrics have also been put forward [44]. Furthermore, a group theory and first-principles computations study has shown that electric DMI exists in ferroelectric materials and exhibits a one-to-one correspondence with its magnetic analogue [25].

These studies imply that electric skyrmions and DWs may well appear in plenty of ferroelectrics as ordinary polar textures.

While polar DW motion triggered by various kinds of stimuli, especially, electric fields, is already demonstrated [21,45–48], realization of driven polar skyrmion motion remains elusive because of the lack of available technological means. In ferroelectrics, there is not an equivalent effect of spin-transfer or spin-orbit torque, and electric dipole waves, an analogue of spin waves, are observed just recently [49] and await thorough exploration of their basic features. Latest investigations [50,51] showed that polar skyrmions can sensitively respond to applied electric fields; typically, they expand or shrink depending on the field polarity as what happens to magnetic skyrmions in magnetic fields [3,22] (Fig. 1). Thus, field-based driving of polar skyrmions represents the most promising route; transplanting the proposed scheme to polar skyrmions seems feasible.

In conclusion, we devise and demonstrate, through micromagnetic simulations, how a magnetic field can move magnetic skyrmions along a nanotrack via a DW and discover equipartition of velocity among the DW and skyrmions. Based on Thiele's analytic model, we reproduce the numerical results and find that the DW–skyrmion and inter-skyrmion repulsive forces play a critical role in driving the skyrmion motion and partitioning the DW velocity. Recognizing recent advances in topological ferroelectrics, we are optimistic that the proposed scheme should also apply to polar skyrmions. Our study is expected to spark research efforts on polar skyrmion dynamics [26].

**Acknowledgements:** X.X. is financially supported by the National Natural Science Foundation of China under Grant No. 11774069 and the Guangdong Provincial Natural Science Foundation of China. Y.Z. acknowledges the support by the Guangdong Special Support Project (Grant No. 2019BT02X030), Shenzhen Fundamental Research Fund (Grant No. JCYJ20210324120213037), Shenzhen Peacock Group Plan (Grant No. KQTD20180413181702403), Pearl River Recruitment Program of Talents (Grant No. 2017GC010293), and National Natural Science Foundation of China (Grant Nos. 11974298 and 61961136006).

**REFERENCES**

[1] N. Nagaosa and Y. Tokura, Topological properties and dynamics of magnetic skyrmions, Nat. Nanotech. 8, 899 (2013).

[2] A. Fert, N. Reyren, and V. Cros, Magnetic skyrmions: advances in physics and potential applications, Nat. Rev. Mater. 2, 17031 (2017).

[3] A. N. Bogdanov and C. Panagopoulos, Physical foundations and basic properties of magnetic skyrmions, Nat. Rev. Phys. 2, 492 (2020).

[4] T. Schulz, R. Ritz, A. Bauer, M. Halder, M. Wagner, C. Franz, C. Pfleiderer, K. Everschor, M. Garst, and A. Rosch, Emergent electrodynamics of skyrmions in a chiral magnet, Nat. Phys. 8, 301 (2012).

[5] S. Parkin and S.-H. Yang, Memory on the racetrack, Nat. Nanotech. 10, 195 (2015).

[6] C. Schütte and M. Garst, Magnon-skyrmion scattering in chiral magnets, Phys. Rev. B 90, 094423 (2014).

[7] J. Iwasaki, A. J. Beekman, and N. Nagaosa, Theory of magnon-skyrmion scattering in chiral magnets, Phys. Rev. B 89, 064412 (2014).

[8] F. Jonietz, S. Mühlbauer, C. Pfleiderer, A. Neubauer, W. Münzer, A. Bauer, T. Adams, R. Georgii, P. Böni, R. A. Duine, K. Everschor, M. Garst, and A. Rosch, Spin transfer torques in MnSi at ultralow current densities, Science 330, 1648 (2010).

[9] L. Kong and J. Zang, Dynamics of an insulating skyrmion under a temperature gradient, Phys. Rev. Lett. 111, 067203 (2013).

[10] M. Mochizuki, X. Z. Yu, S. Seki, N. Kanazawa, W. Koshibae, J. Zang, M. Mostovoy, Y. Tokura, and N. Nagaosa, Thermally driven ratchet motion of a skyrmion microcrystal and topological magnon Hall effect, Nat. Mater. 13, 241 (2014).

[11] S. L. Zhang, W. W. Wang, D. M. Burn, H. Peng, H. Berger, A. Bauer, C. Pfleiderer, G. van der Laan, and T. Hesjedal, Manipulation of skyrmion motion by magnetic field gradients, Nat. Commun. 9, 2115 (2018).

[12] S. Woo, K. Litzius, B. Krüger, M.-Y. Im, L. Caretta, K. Richter, M. Mann, A. Krone, R. M. Reeve, M. Weigand, P. Agrawal, I. Lemesh, M.-A. Mawass, P. Fischer, M. Kläui, and G. S. D. Beach, Observation of room-temperature magnetic skyrmions and their current-driven dynamics in ultrathin metallic ferromagnets, Nat. Mater. 15, 501 (2016).

[13] K. Zeissler, S. Finizio, C. Barton, A. J. Huxtable, J. Massey, J. Raabe, A. V. Sadovnikov, S. A. Nikitov, R. Brearton, T. Hesjedal, G. van der Laan, M. C. Rosamond, E. H. Linfield, G. Burnell, and C. H. Marrows, Diameter-independent skyrmion Hall angle observed in chiral magnetic multilayers, Nat. Commun. 11, 428

(2020).

[14] I. Dzyaloshinsky, A thermodynamic theory of "weak" ferromagnetism of antiferromagnetics, J. Phys. Chem. Solids 4, 241 (1958).

[15] T. Moriya, Anisotropic superexchange interaction and weak ferromagnetism, Phys. Rev. 120, 91 (1960).

[16] S.-Z. Lin, C. Reichhardt, C. D. Batista, and A. Saxena, Particle model for skyrmions in metallic chiral magnets: Dynamics, pinning, and creep, Phys. Rev. B 87, 214419 (2013).

[17] J. Iwasaki, W. Koshibae, and N. Nagaosa, Colossal spin transfer torque effect on skyrmion along the edge, Nano Lett. 14, 4432 (2014).

[18] X. Xing, J. Åkerman, and Y. Zhou, Enhanced skyrmion motion via strip domain wall, Phys. Rev. B 101, 214432 (2020).

[19] M.-W. Yoo, V. Cros, and J.-V. Kim, Current-driven skyrmion expulsion from magnetic nanostrips, Phys. Rev. B 95, 184423 (2017).

[20] O. Boulle, J. Vogel, H. Yang, S. Pizzini, D. de Souza Chaves, A. Locatelli, T. O. Menteş, A. Sala, L. D. Buda-Prejbeanu, O. Klein, M. Belmeguenai, Y. Roussigné, A. Stashkevich, S. M. Chérif, L. Aballe, M. Foerster, M. Chshiev, S. Auffret, I. M. Miron, and G. Gaudin, Room-temperature chiral magnetic skyrmions in ultrathin magnetic nanostructures, Nat. Nanotech. 11, 449 (2016).

[21] Y.-H. Shin, I. Grinberg, I.-W. Chen, and A. M. Rappe, Nucleation and growth mechanism of ferroelectric domain-wall motion, Nature 449, 881 (2007).

[22] N. Romming, A. Kubetzka, C. Hanneken, K. von Bergmann, and R. Wiesendanger, Field-dependent size and shape of single magnetic skyrmions, Phys. Rev. Lett. 114, 177203 (2015).

[23] S. Das, Y. L. Tang, Z. Hong, M. A. P. Gonçalves, M. R. McCarter, C. Klewe, K. X. Nguyen, F. Gómez-Ortiz, P. Shafer, E. Arenholz, V. A. Stoica, S.-L. Hsu, B. Wang, C. Ophus, J. F. Liu, C. T. Nelson, S. Saremi, B. Prasad, A. B. Mei, D. G. Schlom, J. Íñiguez, P. García-Fernández, D. A. Muller, L. Q. Chen, J. Junquera, L. W. Martin, and R. Ramesh, Observation of room-temperature polar skyrmions, Nature 568, 368 (2019).

[24] J. Yin, H. Zong, H. Tao, X. Tao, H. Wu, Y. Zhang, L.-D. Zhao, X. Ding, J. Sun, J. Zhu, J. Wu, and S. J. Pennycook, Nanoscale bubble domains with polar topologies in bulk ferroelectrics, Nat. Commun. 12, 3632 (2021).

[25] H. J. Zhao, P. Chen, S. Prosandeev, S. Artyukhin, and L. Bellaiche, Dzyaloshinskii–Moriya-like interaction in ferroelectrics and antiferroelectrics, Nat. Mater. 20, 341 (2021).

[26] Q. Li, V. A. Stoica, M. Paściak, Y. Zhu, Y. Yuan, T. Yang, M. R. McCarter, S. Das, A. K. Yadav, S. Park, C.

Dai, H. J. Lee, Y. Ahn, S. D. Marks, S. Yu, C. Kadlec, T. Sato, M. C. Hoffmann, M. Chollet, M. E. Kozina, S. Nelson, D. Zhu, D. A. Walko, A. M. Lindenberg, P. G. Evans, L.-Q. Chen, R. Ramesh, L. W. Martin, V. Gopalan, J. W. Freeland, J. Hlinka, and H. Wen, Subterahertz collective dynamics of polar vortices, Nature 592, 376 (2021).

[27] L. Landau and E. Lifshitz, On the theory of the dispersion of magnetic permeability in ferromagnetic bodies, Phys. Z. Sowjetunion 8, 153 (1935).

[28] T. L. Gilbert, A phenomenological theory of damping in ferromagnetic materials, IEEE Trans. Magn. 40, 3443 (2004).

[29] A. Vansteenkiste, J. Leliaert, M. Dvornik, M. Helsen, F. García-Sánchez, and B. Van Waeyenberge, The design and verification of MuMax3, AIP Adv. 4, 107133 (2014).

[30] The exchange length [31] $\lambda = \sqrt{2A/\mu_0 M_s^2} \approx$ 8.4 nm, nominal skyrmion radius [32] $R = \pi D\sqrt{A/(16AK^2 - \pi^2 D^2 K)} \approx$ 5.6 nm, and DW thickness [31] $\Delta = \sqrt{A/K} \approx$ 5.0 nm, for the used typical material parameters.

[31] A. Thiaville and Y. Nakatani, Domain-wall dynamics in nanowires and nanostrips, Spin Dynamics in Confined Magnetic Structures III (Springer, Berlin, 2006), pp. 161–205.

[32] X. S. Wang, H. Y. Yuan, and X. R. Wang, A theory on skyrmion size, Commun. Phys. 1, 31 (2018).

[33] Y. Tokura and N. Kanazawa, Magnetic skyrmion materials, Chem. Rev. 121, 2857 (2021).

[34] S. Rohart and A. Thiaville, Skyrmion confinement in ultrathin film nanostructures in the presence of Dzyaloshinskii-Moriya interaction, Phys. Rev. B 88, 184422 (2013).

[35] G. S. D. Beach, C. Nistor, C. Knutson, M. Tsoi, and J. L. Erskine, Dynamics of field-driven domain-wall propagation in ferromagnetic nanowires, Nat. Mater. 4, 741 (2005).

[36] A. Mougin, M. Cormier, J. P. Adam, P. J. Metaxas, and J. Ferré, Domain wall mobility, stability and Walker breakdown in magnetic nanowires, Europhys. Lett. 78, 57007 (2007).

[37] O. Boulle, S. Rohart, L. D. Buda-Prejbeanu, E. Jué, I. M. Miron, S. Pizzini, J. Vogel, G. Gaudin, and A. Thiaville, Domain wall tilting in the presence of the Dzyaloshinskii-Moriya interaction in out-of-plane magnetized magnetic nanotracks, Phys. Rev. Lett. 111, 217203 (2013).

[38] We would like to note that, for clarity, the cut-off fields on the velocity–field curves in Fig. 2(c) are set to the skyrmion annihilation fields, although, for all ($K_u$, $D$), the DW breakdown fields are concertedly far larger than the skyrmion annihilation fields, as listed in Table S1 [39].

[39] See Supplemental Material at URL for Walker breakdown of a DW, skyrmion annihilation in the DW–

skyrmion motion, and their corresponding threshold fields (Section A) as well as for evidence of the general validity of the presented results (Section B).

[40] O. Tchernyshyov and G.-W. Chern, Fractional vortices and composite domain walls in flat nanomagnets, Phys. Rev. Lett. 95, 197204 (2005).

[41] Y. Nakatani, A. Thiaville, and J. Miltat, Faster magnetic walls in rough wires, Nat. Mater. 2, 521 (2003).

[42] A. A. Thiele, Steady-state motion of magnetic domains, Phys. Rev. Lett. 30, 230 (1973).

[43] A. Hrabec, J. Sampaio, M. Belmeguenai, I. Gross, R. Weil, S. M. Chérif, A. Stashkevich, V. Jacques, A. Thiaville, and S. Rohart, Current-induced skyrmion generation and dynamics in symmetric bilayers, Nat. Commun. 8, 15765 (2017).

[44] M. A. P. Gonçalves, C. Escorihuela-Sayalero, P. García-Fernández, J. Junquera, and J. Íñiguez, Theoretical guidelines to create and tune electric skyrmion bubbles, Sci. Adv. 5, eaau7023 (2019).

[45] T. J. Yang, V. Gopalan, P. J. Swart, and U. Mohideen, Direct observation of pinning and bowing of a single ferroelectric domain wall, Phys. Rev. Lett. 82, 4106 (1999).

[46] L. J. McGilly, P. Yudin, L. Feigl, A. K. Tagantsev, and N. Setter, Controlling domain wall motion in ferroelectric thin films, Nat. Nanotech. 10, 145 (2015).

[47] J. R. Whyte and J. M. Gregg, A diode for ferroelectric domain-wall motion, Nat. Commun. 6, 7361 (2015).

[48] F. Rubio-Marcos, A. D. Campo, P. Marchet, and J. F. Fernández, Ferroelectric domain wall motion induced by polarized light, Nat. Commun. 6, 6594 (2015).

[49] F. H. Gong, Y. L. Tang, Y. L. Zhu, H. Zhang, Y. J. Wang, Y. T. Chen, Y. P. Feng, M. J. Zou, B. Wu, W. R. Geng, Y. Cao, and X. L. Ma, Atomic mapping of periodic dipole waves in ferroelectric oxide, Sci. Adv. 7, eabg5503 (2021).

[50] S. Das, Z. Hong, V. A. Stoica, M. A. P. Gonçalves, Y. T. Shao, E. Parsonnet, E. J. Marksz, S. Saremi, M. R. McCarter, A. Reynoso, C. J. Long, A. M. Hagerstrom, D. Meyers, V. Ravi, B. Prasad, H. Zhou, Z. Zhang, H. Wen, F. Gómez-Ortiz, P. García-Fernández, J. Bokor, J. Íñiguez, J. W. Freeland, N. D. Orloff, J. Junquera, L. Q. Chen, S. Salahuddin, D. A. Muller, L. W. Martin, and R. Ramesh, Local negative permittivity and topological phase transition in polar skyrmions, Nat. Mater. 20, 194 (2021).

[51] Y. Nahas, S. Prokhorenko, Q. Zhang, V. Govinden, N. Valanoor, and L. Bellaiche, Topology and control of self-assembled domain patterns in low-dimensional ferroelectrics, Nat. Commun. 11, 5779 (2020).

**FIGURE CAPTIONS**

**FIG. 1.** Isolated skyrmion's response to a perpendicular magnetic field. (a) (b) Equilibrium spin configurations of an isolated skyrmion under different magnetic fields in a magnetic nanotrack (60 × 2000 nm$^2$) and a square element (300 × 300 nm$^2$). Field strength is indicated on each panel. Determined by the field polarity, the skyrmion either expands or contracts without a net displacement. The color scale for $m_z$ is used throughout this paper. (c) Skyrmion size as function of field strength. Strong edge confinement in the nanotrack suppresses inflation in skyrmion area.

**FIG. 2.** Field-driven dynamics of a single DW and a DW–skyrmion pair. (a) DW and (b) DW–skyrmion pair motion in a positive field ($H_z$ = 400 Oe) through expansion of the up domain. The time after applying the field is indicated on each panel. Through a similar distance ($\Delta x$: 209 vs 203 nm), the DW–skyrmion pair spends twice as much time as the single DW ($\Delta t$: 3.4 vs 1.7 ns), meaning that the DW speed roughly halves when a skyrmion is carried. (c) DW and DW–skyrmion velocities against field for various parameters as indicated. Irrespective of the field strength and material parameters, the velocities of the DW–skyrmion pair are always nearly half of those of the single DW, suggesting the universality of the feature.

**FIG. 3.** Field-driven dynamics of a DW–skyrmion train. (a) Migration of the DW–skyrmion ensemble as a whole accompanying expansion of the up domain in a positive field ($H_z$ = 400 Oe). Here, the train consists of 5 skyrmions. The time after applying the field is indicated on each panel. Over the latter two durations ($\Delta t$: 8 ns, i.e., 8–16 and 16–24 ns), the ensemble travels equal distances ($\Delta x$: 174 nm), implying that at $t$ = 8 ns the steady state has been attained. Consequently, the ensemble velocity can be calculated as $v = \Delta x / \Delta t$. (b) Ensemble velocity against magnetic field for $N_s$ = 0, 1, 2, 3, 4, and 5.

Definitely, an increase in the number of skyrmions leads to a reduction in the ensemble velocity. (c) Skyrmion diameters and spacings under different field strengths for $N_s = 5$. $d_{SKi}$ denotes the diameter of the $i^{th}$ skyrmion away from the DW. $S_{DW-SK1}$ and $S_{SKi-(i+1)}$ signify the spacing between the DW and $1^{st}$ skyrmion and that between the $i^{th}$ and $(i+1)^{th}$ skyrmions.

**FIG. 4.** Ensemble velocity evolution with the number of included skyrmions. (a) Ensemble velocity versus $N_s$ under various field strengths. Lines serve as guides to eyes. (b) Ensemble mobility versus $N_s$. Black dots are simulation results and red dots are derived from $\tilde{u}_{dw} / (N_s + 1)$ with $\tilde{u}_{dw}$ representing the mobility of a single DW.

**FIG. 5.** Thiele forces acting on the DW and skyrmions in steady motion. Here, $N_s = 3$ is taken as an example and the transverse forces are ignored.

**FIG. 6.** Selective delivery of a skyrmion via a DW in terms of the field polarity. In such configuration, a DW sits in between two sets of skyrmions and a given field will shift only one set of skyrmions in a certain direction without affecting the other set. Note that, more than one skyrmion can be included in each set.

# Figure 1

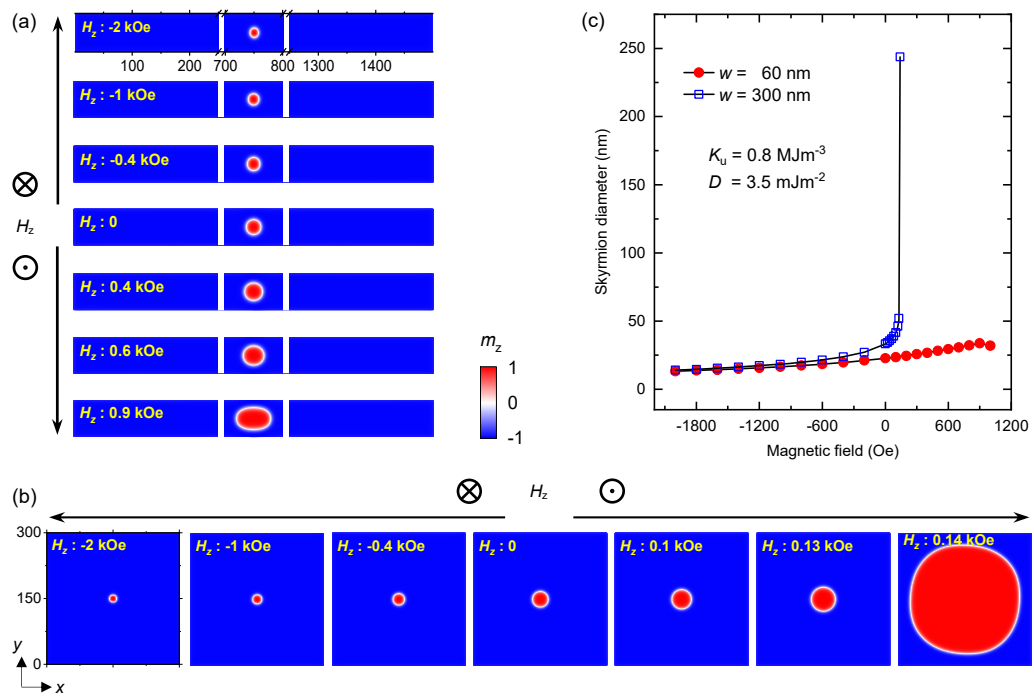

**Figure 2**

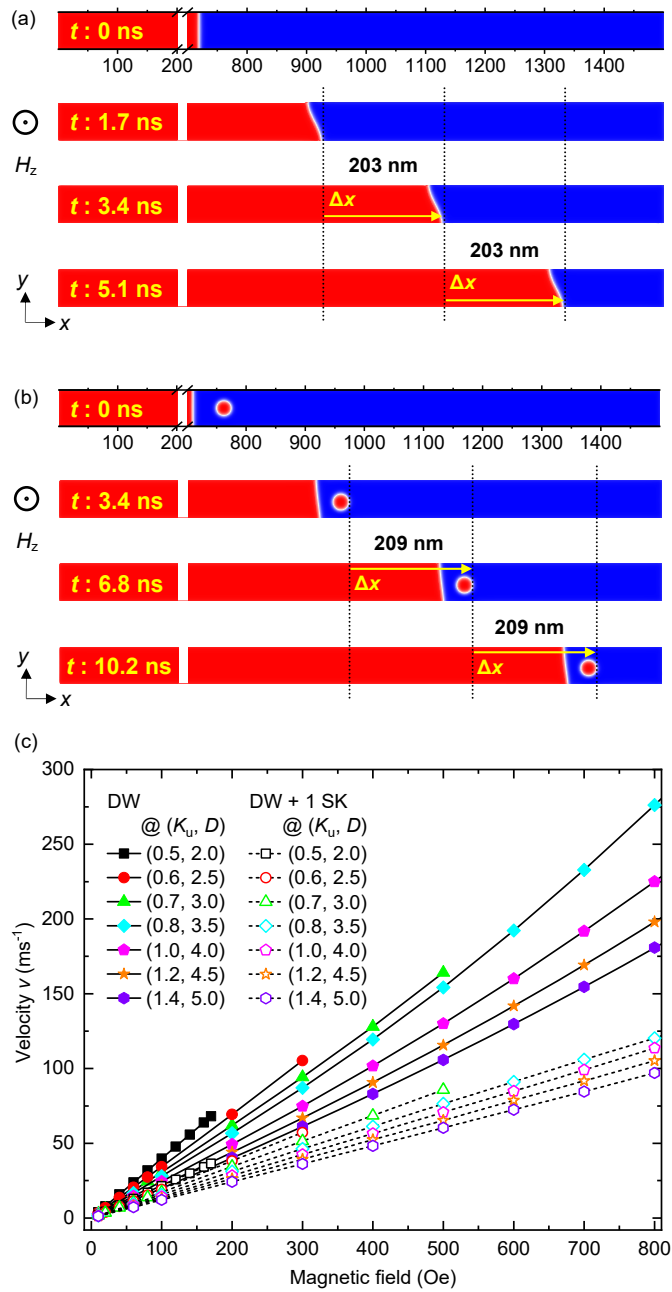

**Figure 3**

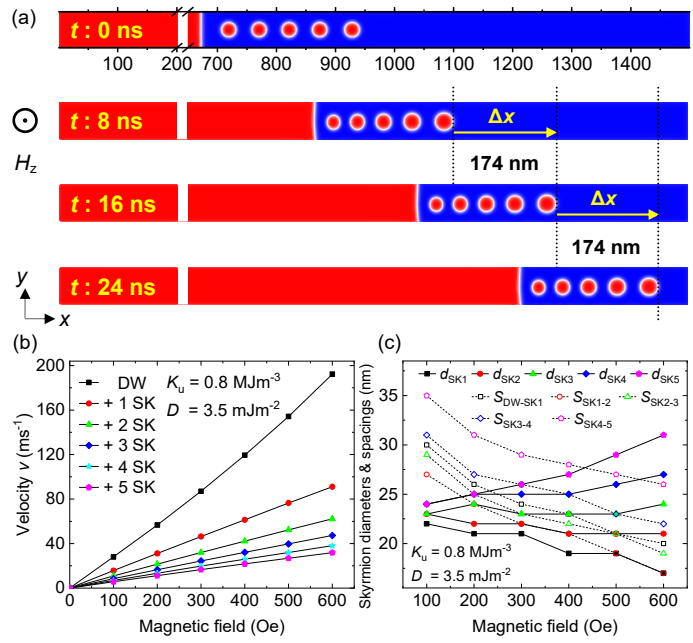

**Figure 4**

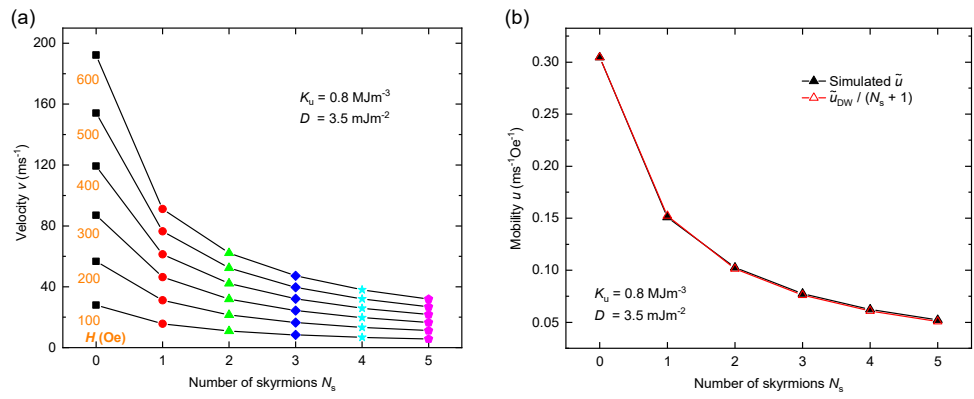

**Figure 5**

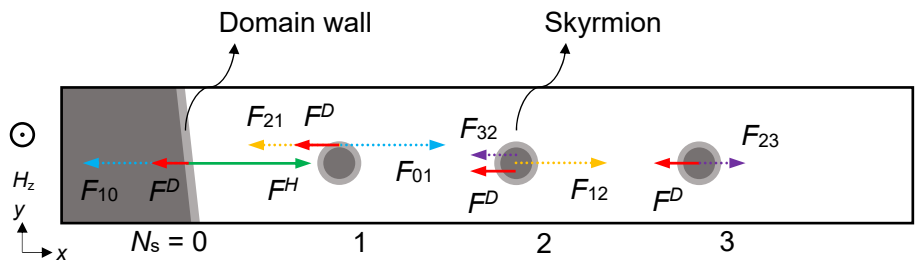

**Figure 6**

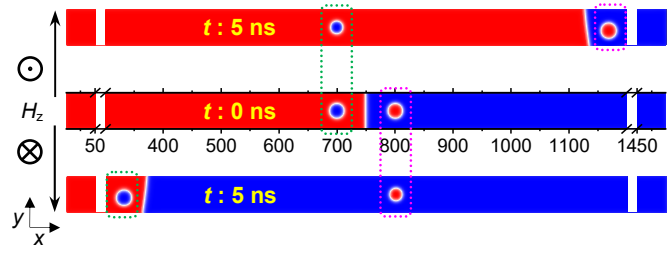

# Supplemental Material for

# Field-driven skyrmion motion through velocity equipartition between skyrmions and a domain wall

Xiangjun Xing[1,2],* and Yan Zhou[3],*

[1]*School of Physics & Optoelectronic Engineering, Guangdong University of Technology, Guangzhou 510006, China*

[2]*Guangdong Provincial Key Laboratory of Information Photonics Technology, Guangdong University of Technology, Guangzhou 510006, China*

[3]*School of Science & Engineering, The Chinese University of Hong Kong, Shenzhen, Guangdong 518172, China*

**This PDF file includes:**

---

*Email: xjxing@gdut.edu.cn; zhouyan@cuhk.edu.cn

# SECTION A

**TABLE SI.** Threshold magnetic fields ($H_{th}$) associated with the motions of a single DW and a DW-mediated skyrmion (DWmSK). Above the threshold fields, neither the single DW nor the DW-mediated skyrmion can maintain steady motion. For the DW motion, the threshold fields correspond to initiation of Walker breakdown as shown in Fig. S1, whereas for the DW-mediated skyrmion motion, they signify onset of skyrmion annihilation as shown in Fig. S2. Various combinations of ($K_u$, $D$) are considered with $K_u$ and $D$ being in units of MJm$^{-3}$ and mJm$^{-2}$, respectively.

| $H_{th}$ (Oe) | ($K_u$, $D$) | | | | | | | | |
|---|---|---|---|---|---|---|---|---|---|
| | (0.5, 2.0) | (0.6, 2.5) | (0.7, 3.0) | (0.8, 3.0) | (0.8, 3.5) | (1.0, 4.0) | (1.2, 4.5) | (1.3, 4.5) | (1.4, 5.0) |
| DW | 2100 | 2900 | 3800 | 4200 | 4900 (5000)* | 6500 | 8200 | 8500 | 10000 |
| DWmSK | 170 | 300 | 500 | 390 | 850 (940)* | 950 | 1140 | 940 | 1300 |

Data with and without asterisk correspond to Bloch-type (bulk DMI) and Néel-type (interfacial DMI) spin textures, respectively.

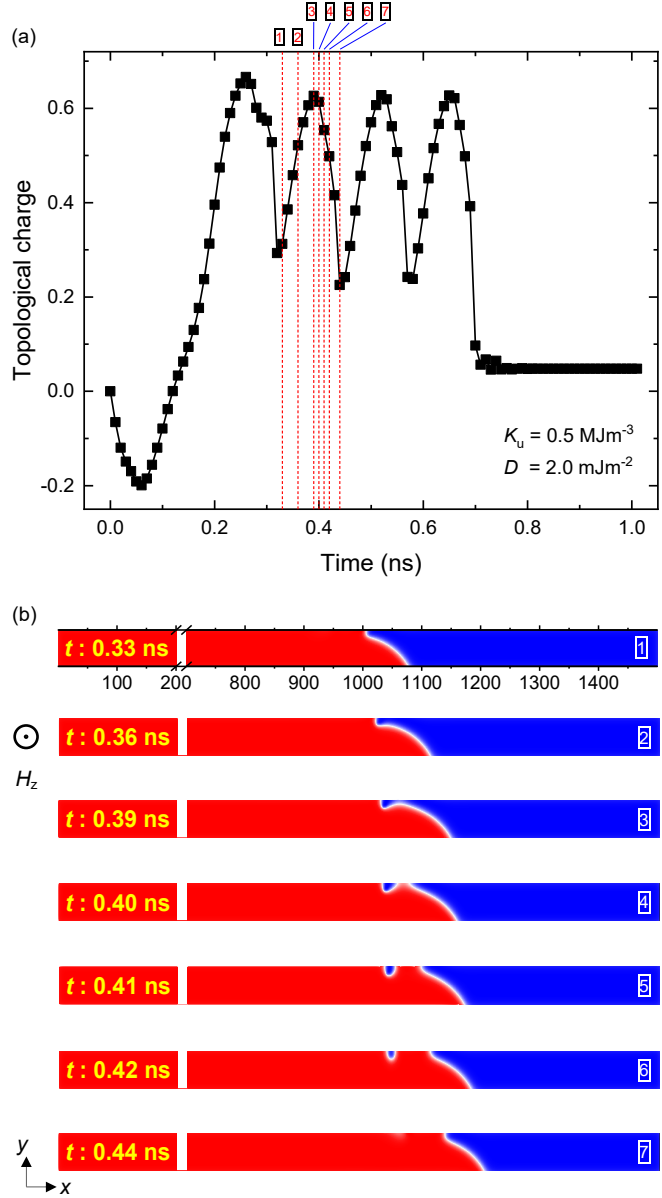

**FIG. S1. DW dynamics with Walker breakdown.** (a) Evolution of the system's total topological charge during DW motion at $H_z$ = 2.4 kOe. Periodic oscillations on the curve feature the Walker breakdown process. Each cycle involves several typical events labeled as 1–7. (b) DW structure transformation corresponding to the Walker breakdown, incorporating antivortex nucleation around the edge (1), growth (2–3), division from the DW (4), shrinkage (4–6), as well as annihilation at the edge (6–7) on a short timescale of ~ 0.1 ns. This process occurs repeatedly until the DW touches the far edge.

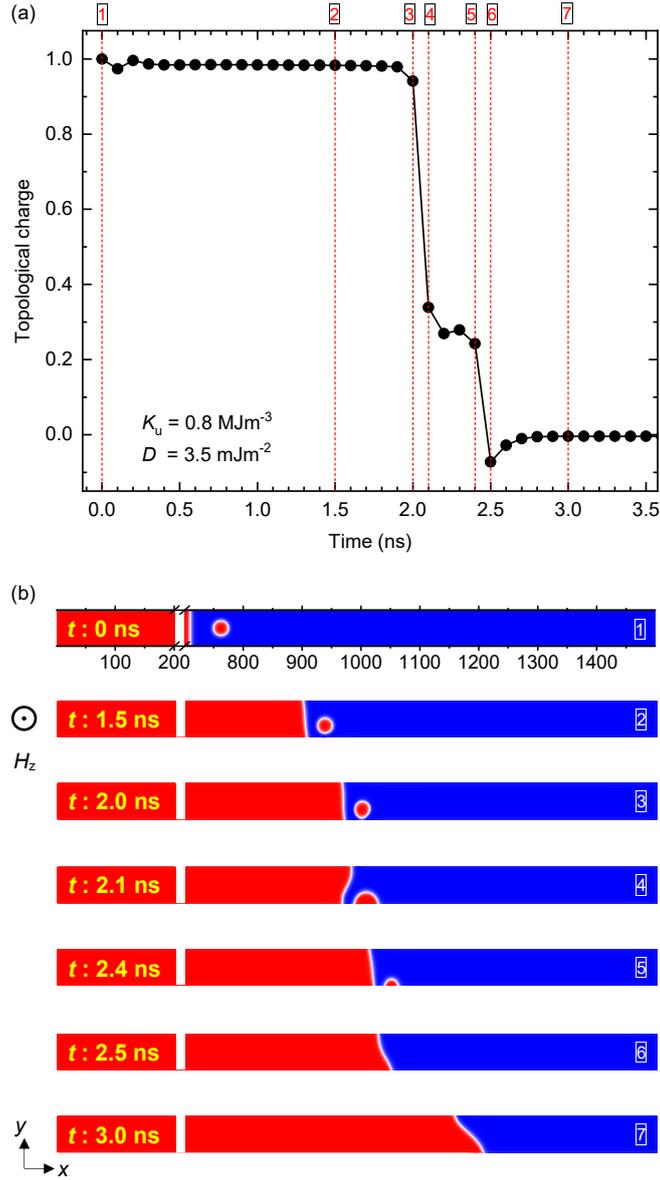

**FIG. S2. Skyrmion annihilation in field-driven DW–skyrmion motion.** (a) Topological charge evolution during DW–skyrmion motion at $H_z$ = 880 Oe. The steep drop in topological charge from 1 to 0 (3–6) characterizes skyrmion annihilation. (b) Skyrmion structure destruction (3–4) and annihilation (5–6) at the edge.

# SECTION B

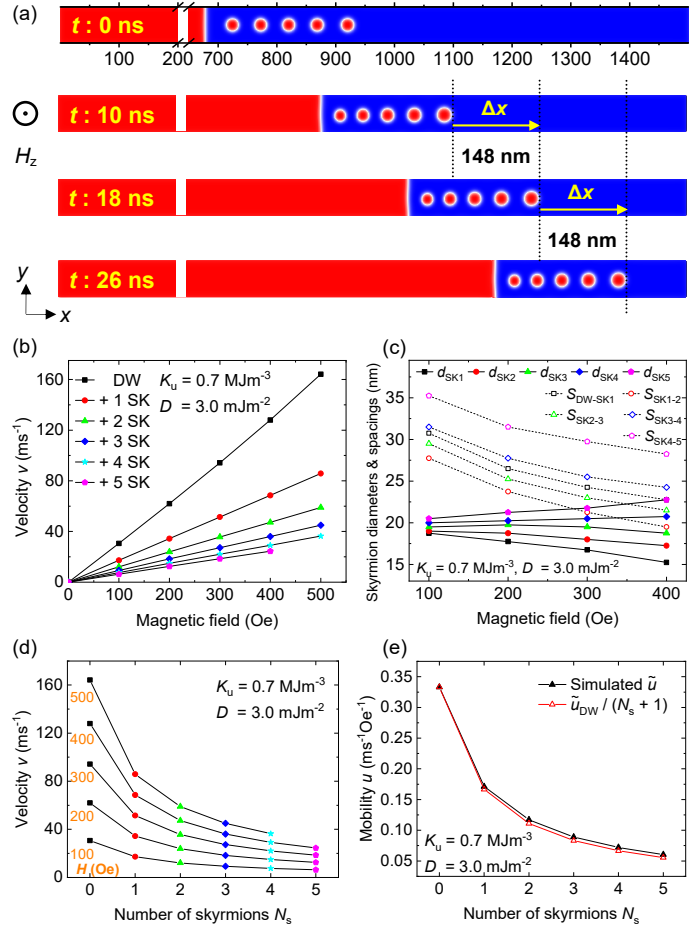

**FIG. S3. Field-driven dynamics of DW–skyrmion train for a different set of $K_u$ and $D$.** (a) Migration of the DW–skyrmion ensemble as a whole accompanying expansion of the up domain in a positive field ($H_z$ = 300 Oe). Here, the train consists of 5 skyrmions. $K_u$ = 0.7 MJm$^{-3}$ and $D$ = 3.0 mJm$^{-2}$. The time after applying the field is indicated on each panel. Over the latter two durations ($\Delta t$: 8 ns, i.e., 10–18 and 18–26 ns), the ensemble travels equal distances ($\Delta x$: 148 nm), implying that at $t$ = 10 ns the steady state has been attained. Accordingly, the ensemble velocity can be calculated as $v = \Delta x / \Delta t$. (b) Ensemble velocity against magnetic field for $N_s$ = 0, 1, 2, 3, 4, and 5. Definitely, an increase in the number of skyrmions leads to a reduction in the ensemble velocity. (c) Skyrmion diameters and spacings under different field strengths for $N_s$ = 5. $d_{SKi}$ denotes the diameter of the $i^{th}$ skyrmion away from the DW. $S_{DW–SK1}$ and $S_{SKi–(i+1)}$ signify the spacing between the DW and 1$^{st}$ skyrmion and that between the $i^{th}$ and $(i+1)^{th}$ skyrmions. (d) Ensemble velocity evolution with $N_s$ under various field strengths. Lines are guides to eyes. (e) Ensemble mobility versus $N_s$. Black dots are simulation results and red dots are derived from $\tilde{u}_{dw} / (N_s + 1)$ with $\tilde{u}_{dw}$ representing the mobility of a single DW.

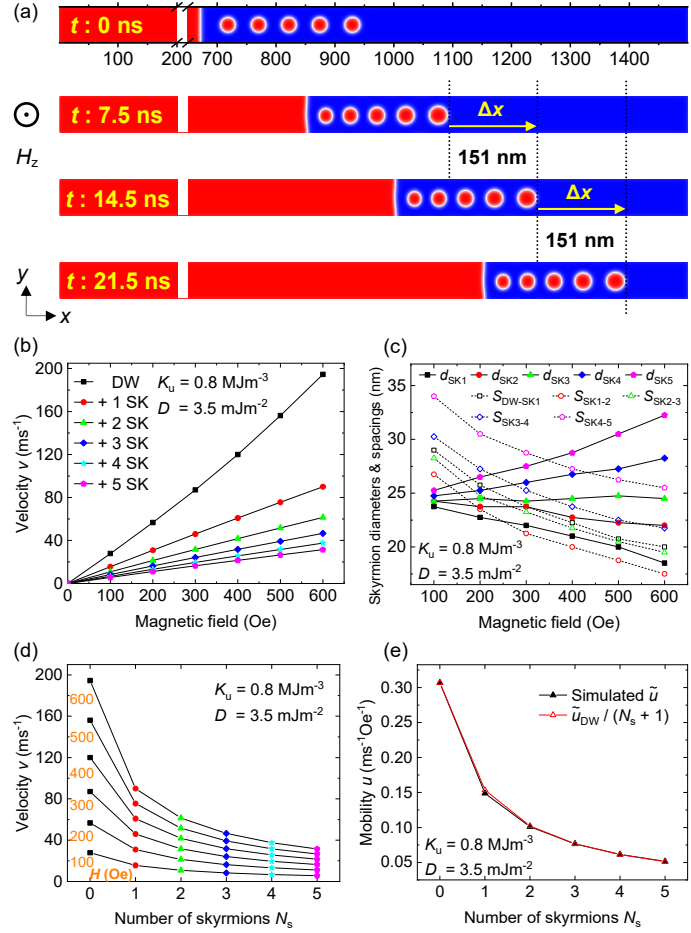

**FIG. S4. Field-driven dynamics of Bloch-type DW–skyrmion train.** (a) Migration of the DW–skyrmion ensemble as a whole accompanying expansion of the up domain in a positive field ($H_z$ = 400 Oe). Here, the train consists of 5 skyrmions. $K_u$ = 0.8 MJm$^{-3}$ and $D$ = 3.5 mJm$^{-2}$. Bloch characteristic of the DW and skyrmions is rendered by using a bulk DMI. The time after applying the field is indicated on each panel. Over the latter two durations ($\Delta t$: 7 ns, i.e., 7.5–14.5 and 14.5–21.5 ns), the ensemble travels equal distances ($\Delta x$: 151 nm), implying that at $t$ = 7.5 ns the steady state has been attained. Accordingly, the ensemble velocity can be calculated as $v = \Delta x / \Delta t$. (b) Ensemble velocity against magnetic field for $N_s$ = 0, 1, 2, 3, 4, and 5. Definitely, an increase in the number of skyrmions leads to a reduction in the ensemble velocity. (c) Skyrmion diameters and spacings under different field strengths for $N_s$ = 5. $d_{SKi}$ denotes the diameter of the $i^{th}$ skyrmion away from the DW. $S_{DW–SK1}$ and $S_{SKi–(i+1)}$ signify the spacing between the DW and 1$^{st}$ skyrmion and that between the $i^{th}$ and $(i+1)^{th}$ skyrmions. (d) Ensemble velocity evolution with $N_s$ under various field strengths. Lines are guides to eyes. (e) Ensemble mobility versus $N_s$. Black dots are simulation results and red dots are derived from $\tilde{u}_{dw} / (N_s + 1)$ with $\tilde{u}_{dw}$ representing the mobility of a single DW.